\documentclass{article}

	\usepackage[usenames,dvipsnames,svgnames,x11names]{xcolor}
	
	\newcommand{\Note}[1]{\textcolor{red}{#1}}
	\newcommand{\ysnoted}[1]{}
	
	\makeatletter
	\def\mathcolor#1#{\@mathcolor{#1}}
	\def\@mathcolor#1#2#3{%
	  \protect\leavevmode
	  \begingroup
		\color#1{#2}#3%
	  \endgroup
	}
	\makeatother
	
	\usepackage[normalem]{ulem}

	\usepackage[nocompress]{cite}

	\usepackage{listings}
	
	\lstdefinelanguage{XML}
	{
	basicstyle=\ttfamily\footnotesize,
	  morestring=[b]",
	  moredelim=[s][\bfseries\color{Maroon}]{<}{\ },
	  moredelim=[s][\bfseries\color{Maroon}]{</}{>},
	  moredelim=[l][\bfseries\color{Maroon}]{/>},
	  moredelim=[l][\bfseries\color{Maroon}]{>},
	  morecomment=[s]{<?}{?>},
	  morecomment=[s]{<!--}{-->},
	  commentstyle=\color{gray},
	  stringstyle=\color{blue},
	  identifierstyle=\color{red}
	}

	\usepackage{moreverb}
	
	\usepackage[nounderscore]{syntax}
	
	\usepackage[pdftex]{graphicx}
	\usepackage{subfig}
	
	\graphicspath{{./figures/}}

	\DeclareGraphicsExtensions{.pdf}

	\usepackage{subfig}

	\usepackage[cmex10]{amsmath}
	\usepackage{amssymb}
	\usepackage{mathtools}
	\usepackage{amsthm}
	\usepackage{amsfonts}
	\usepackage{bbm}

	\usepackage{algorithmicx}
	\usepackage[noend]{algpseudocode}
	\usepackage[ruled]{algorithm}
	\definecolor{light-gray}{gray}{0.75}
	\algrenewcommand{\algorithmiccomment}[1]{\hskip3em{{\footnotesize \textcolor{light-gray}{$\blacktriangleright$}}} #1}

	\usepackage{multirow} 
	\usepackage{rotating} 
	\usepackage{booktabs} 
	\usepackage{colortbl} 
	\usepackage{tablefootnote} 
	
	\usepackage[pdftex,colorlinks=true,urlcolor=blue,citecolor=blue]{hyperref}
	
	\usepackage{xspace}

	\usepackage{enumitem}

	\usepackage[htt]{hyphenat}

	\newcommand{\elf}{ElfStore\xspace}

	\usepackage[english]{babel}
	\usepackage{blindtext}

	\date{}
	
\begin{document}
	\title{\emph{\elf}: A Resilient Data Storage Service for Federated Edge and Fog Resources \footnote{To appear in IEEE International Conference on Web Services (ICWS), Milan, Italy, 2019}}
			
	\author{ Sumit Kumar Monga, Sheshadri K R and Yogesh Simmhan \\
		\emph{Department of Computational and Data Sciences,}\\
		\emph{Indian Institute of Science (IISc), Bangalore 560012, India}\\  
		\emph{Email: sumitkm@iisc.ac.in, sheshadrik@iisc.ac.in, simmhan@iisc.ac.in}}
		
\maketitle
	
	\begin{abstract}
Edge and fog computing have grown popular as IoT deployments become wide-spread. While application composition and scheduling on such resources are being explored, there exists a gap in a distributed data storage service on the edge and fog layer, instead depending solely on the cloud for data persistence. Such a service should reliably store and manage data on fog and edge devices, even in the presence of failures, and offer transparent discovery and access to data for use by edge computing applications. Here, we present \emph{\elf}, a first-of-its-kind \underline{e}dge-\underline{l}ocal \underline{f}ederated \underline{store} for streams of data blocks. It uses reliable fog devices as a super-peer overlay to monitor the edge resources, offers federated metadata indexing using Bloom filters, locates data within 2-hops, and maintains approximate global statistics about the reliability and storage capacity of edges. Edges host the actual data blocks, and we use a unique differential replication scheme to select edges on which to replicate blocks, to guarantee a minimum reliability and to balance storage utilization. Our experiments on two IoT virtual deployments with $20$ and $272$ devices show that \elf has low overheads, is bound only by the network bandwidth, has scalable performance, and offers tunable resilience.
	\end{abstract}

	\section{Introduction}
	\label{sec:intro}
The growing prevalence of \emph{Internet of Things (IoT)} deployments as part of smart city and industrial infrastructure is leading to a rapid influx of \emph{data generated continuously} from thousands of sensors~\cite{perera2014}. These data sources include smart utility meters, air pollution monitors, security cameras, and equipment sensors. \emph{Analytics} over these data, in real-time or periodically, helps make intelligent decisions for the efficient and reliable management of such complex systems~\cite{simmhan2013}. 

At the same time, IoT is also leading to the availability of \emph{edge and fog computing devices} on the field, as part of sensors and gateways~\cite{dastjerdi2016}. Affordable edge devices like Raspberry Pi are often co-located with the sensors on private and wide-area networks to \emph{acquire} data, perform \emph{local analytics}, and \emph{transmit} it to cloud data centers for persistence~\cite{xu2017}. Fog devices like NVidia Jetson TX2 \emph{manage} neighboring edge devices on the network, offer more \emph{advanced computing} for further analytics or aggregation, and also \emph{forward} data to the cloud. In large IoT deployments, the edge and fog devices are often organized in a \emph{2-level hierarchy} for ease of management and scalability~\cite{he2017}, and complemented by cloud resources.

\emph{Edge computing} is motivated by the access to such cheap or free edge and fog compute resources, the reduced network latency between the data source and the analytics that makes the decision (e.g., power grid management), and to mitigate network use by high-bandwidth applications (e.g., video analytics for urban safety)~\cite{garcia2015,yannuzzi2017}. There is active research on composing micro-services and scheduling dataflows for execution on edge and fog resources, in combination with or instead of cloud resources~\cite{ravindra2017,wu2018}. These \emph{platform services} allow applications to run continuously over incremental data.

However, two key gaps exist. One, there is a lack of transparent \emph{data access service} at the edge or fog, from which such applications can consume their input. Typically, streaming application bind to specific device endpoints or topics on a central publish-subscribe broker, while file-based applications use \emph{ad hoc} mechanisms. Ideally, applications should be able to use the logical features of the data they are interested in, such as its metadata, rather than its physical address, to access it. Two, data generated on the edge and fog are only transiently available on them, and eventually \emph{moved to the cloud} for persistence, a key reason being that edge devices are usually less reliable. So, applications using such data are forced to run on the cloud, or move them back to the edge for computing.

These motivate the need for a \emph{distributed data storage and management service} over fog and unreliable edge devices that offers \emph{content-based discovery, transparent access, and high availability of data}, across a wide area network and in the presence of device failures. This ensures data locality for application micro-services on the edge, allows the cumulative storage capacity of the edge devices to be efficiently used, and avoids transferring data to the cloud for persistence. The storage service should also be optimized for data that is \emph{continuously generated}, as is common for IoT sensor data, and yet allow access to different temporal or logical segments within the data stream.

We make the following specific contributions in this paper:
\begin{enumerate}
\item We propose \emph{\elf, an \underline{E}dge-\underline{l}ocal \underline{f}ederated \underline{Store}}, which is a \textbf{first-of-its-kind} \emph{stream-based, block-oriented distributed storage service} over unreliable edge devices, with fog devices managing the operations using a \emph{super-peer overlay network}.

\item We propose a \emph{federated indexing model} using Bloom filters maintained by fogs for a \emph{scalable, probabilistic search} for blocks based on their metadata properties. 
\item We offer tunable resilience for blocks using a \emph{novel differential replication scheme} across unreliable edges. This uses \emph{approximate global statistics} at the fogs to decide on replica placement, which is sensitive to edge reliability, balances capacity usage, and ensures data durability.
\end{enumerate}

The rest of the paper is organized as follows. We review related work to highlight the novelty of our contributions in Sec.~\ref{sec:related}, introduce the \elf service architecture and operations, federated indexing and tunable replication in Sec.~\ref{sec:arch}, present detailed experiments to validate the design and scalability in Sec.~\ref{sec:results}, and offer our conclusions in Sec.~\ref{sec:conclusions}.
	
\section{Related Work}
\label{sec:related}
	
There has been limited work on distributed data storage on edge and fog resources, as reviewed and classified in \emph{Moyasiadis, et al.}~\cite{moysiadis2018}. Rather than off-load to cloud or aggregate to reduce the size, we instead adopt a peer-to-peer (P2P) model which does not reduce data fidelity, and maintains locality on edge and fog resources, with reliability guarantees. Others~\cite{confais2017a} have evaluated existing distributed cloud object stores, \emph{Rados (Ceph), Cassandra} and \emph{Inter Planetary File System (IPFS)}, for use on edge and fog resources, and proposed extensions. However, these store data only on the fog layer, with the fog assumed to be high-end Xeon servers with $128$~GB RAM. We instead design our storage service for practical and large-scale edge and fog resources that run on Pi- and Jetson-class devices with $4$--$8$ ARM cores and $1$--$2$~GB RAM, and use the edge devices as first-class entities for persistence. 

\emph{IPFS}~\cite{benet2014} is used for storing web content on a wide-area network. It uses a Merkle tree to capture the directory structure, content-based addressing for files, 
and a P2P Distributed Hash Table (DHT) to map the file's hash to its peer locations. BitTorrent is used for data movement, and the data is replicated when a client downloads it. \emph{Confais, et al.}~\cite{confais2017} have deployed IPFS on fog and cloud resources using Network Attached Storage (NAS). They extend IPFS to support searching at the local fog, besides the DHT, to speed up access to local content. However, storage is limited to the fog and not edge, and there is no active replication to ensure reliability upon failures.

\emph{FogStore}~\cite{mayer2017} proposes a distributed key-value store on fog resources with replication and differential consistency. Our focus is on reliably storing a stream of blocks of a much larger size, where resilience and capacity constraints are met.
Others~\cite{psaras2018} propose repositories hosted on stable 
fogs (referred to as ``edges'') that are populated by data from transient edges (``mobile devices''), and act as a reverse-Content Distribution Network (CDN) to serve requests from the cloud too. Reliability is a non-goal in their design and no experiments are presented.
\emph{vStore}~\cite{gedeon2018} supports context-aware placement of data on fog and cloud resources, with mobile devices generating and consuming these data. It uses a rules engine to place and locate data based on its context metadata, but ignores reliability as edge devices do not store data.

\emph{Chen, et al.}~\cite{chen2015} examine fault-tolerant and energy-efficient data storage and computation on a set of edge devices (``mobile clouds''), without any fog or cloud. They use \emph{k-of-n erasure coding}, where files are fragmented and 
coded fragments placed on energy-efficient edge devices.
Access to data is by creating \emph{n} tasks that execute on the edge devices containing the fragments, and waiting for \emph{k} of them to complete, so as to decode and process the original fragment. This tightly-couples processing with storage on the same devices, rather than offer an independent data service like us. Also, it is designed for $10$--$100$'s of edge devices since all-to-all information is required for decision making, while we use fog overlays that can scale to $100$'s of fogs and $1000$'s of edges. They do not support searching by metadata like we do. Lastly, erasure codes while space-efficient compared to replication, are time-inefficient for recovery on unreliable systems, like the ones we consider~\cite{plank2013}.

\emph{RFS}~\cite{dong2011} is a distributed file system hosted on the cloud but optimized for mobile clients (edges) with transient network connectivity. While the cloud holds the encrypted master data, clients selectively pre-fetch, decrypt and cache parts of the file based on their access patterns. Clients have exclusive access to their encrypted home directory, and common access to shared directories. The master data in the cloud is reliable.

\emph{P2P systems} like Chord, Pastry and BitTorrent have proposed distributed file, block and key-value storage on unreliable peers on wide-area networks~\cite{stoica2002}. We adopt several of these concepts such as super-peers~\cite{yang2003}, but simplify and enhance their performance for edge and fog deployments with less device flux, guarantee a minimum durability for stored blocks, and balance the storage capacity across peers. We also use an efficient federated indexing using Bloom filters~\cite{ledlie2002}.

\emph{Cloud storage services} like \emph{HDFS} and \emph{Ceph}~\cite{ceph} have been vital to the success of Big Data platforms by separating the distributed storage layer from the computing layer, like \emph{Apache Spark} or \emph{MapReduce}, while allowing co-location during scheduling. We adopt a similar model for edge and fog, while being aware of the network topology, sensitive to variable failure rates of edges, and offering search capability.	

In summary, none of the existing literature or systems provide a \emph{scalable distributed store} for \emph{storing, searching and accessing streams of objects} generated from IoT sensing devices on \emph{fog and unreliable edges}, while \emph{guaranteeing reliability}, \emph{balancing capacity}, and leveraging the \emph{topology} of fog and edge resources.
	
	\section{\elf Architecture}
	\label{sec:arch}

In this section, we describe the desiderata, the supported operations, our design choices, and the architecture for \emph{\underline{E}dge-\underline{l}ocal \underline{f}ederated \underline{Store} (\elf)}.

Our \textbf{system model} has two types of resources, \emph{edge} and \emph{fog}. Edges like Raspberry Pi have constrained compute and memory (e.g., $4$-core ARM32 CPU, $1$~GB RAM), and about $64$~GB of SD card storage.  These commodity devices are cheap but \emph{unreliable}, especially when operating in the field, and have an expected failure rate. Each edge \emph{connects to a single fog}, through a wireless or wired \emph{private local area network} (W/LAN), and the fog manages it. Fogs like Jetson TX2 have moderate resource capacity (e.g., $8$-core ARM64 CPU, $4$~GB RAM, $500$~GB HDD), and serve as a \emph{gateway to the public Internet} for their edges to connect to other fogs and their edges. Fog resources are \emph{reliable}, and connect with each other through a wired Metropolitan or Wide Area Network (MAN/WAN). We plan to support \emph{city-scale deployments} having $10$--$100$'s of fogs, each managing $10$--$100$'s of edges~\cite{yannuzzi2017}.

Given this, there are several \textbf{design goals} and assumptions for our data storage service.
(1) Applications running on edge, fog or other devices on the Internet may \emph{put, search and get data and associated metadata} from the service. However, we expect that the edges will be the \emph{predominant clients} to the store, generating and writing data continuously from co-located sensors, and consuming data for edge micro-services. 
(2) The edges will serve as the \emph{primary storage hosts} for the data to enhance locality (hence, ``edge-local''), with the fogs used for \emph{management and discovery}. We avoid cloud as a storage location, though it can have clients that access the data for processing or long-term archival.
(3) Data that is stored must meet a minimum \emph{reliability} level, even with edge failures, and have sufficient \emph{availability}. The typical \emph{lifetime} of the hosted data is in days or months (not years), as edge applications are likely to be interested in recent data. Adequate \emph{cumulative storage capacity} should be available on the edges.
(4) The store should \emph{scale} as edges join and leave the system, often triggered by device failures and their stateless recovery, or occasional capacity expansion. Its \emph{performance} should also weakly scale with the number of clients.
(5) We assume a \emph{fully-trusted environment}, where all edge and fog devices are secure, part of the same management domain, and there are no access restrictions to the contents.

\begin{figure}[t]
\centering
{
     \includegraphics[width=0.5\columnwidth]{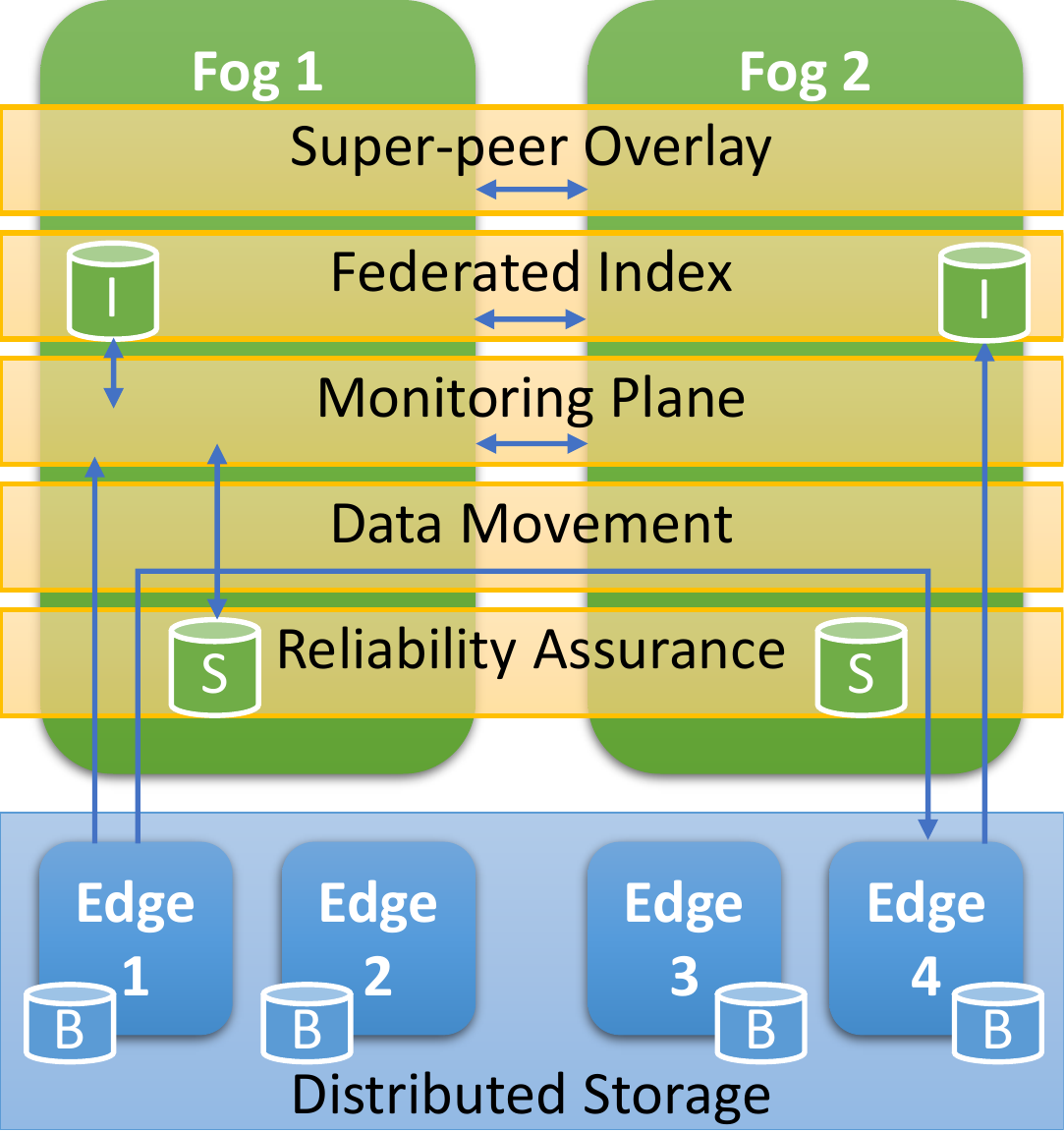}

\caption{High-level Architecture of \elf}
     \label{fig:arch}
     
     \vspace{-0.15in}
}
\end{figure}

The \textbf{\elf} architecture (Fig.~\ref{fig:arch}) addresses these requirements, and offers a federated storage service for streams of blocks. It uses the local disks on unreliable edges in the LAN as the \emph{persistent layer}, and fogs on the WAN connected using a \emph{super-peer overlay} as the \emph{management layer}. It guarantees reliability at the block-level using \emph{differential replication}, and helps search for streams and blocks over their metadata using federated \emph{Bloom filter} indexes. These are discussed next.

\subsection{Data Model and Operations}
IoT data is often streaming, and arrives continuously from sensors. While publish-subscribe brokers enable access for real-time processing, we handle data storage and application access in the short- and medium-term. Since this data accumulates over time, \elf adopts a hybrid data model consisting of a \emph{stream of blocks}. Here, the storage namespace has a flat set of streams, identified by unique stream IDs, and a sequence of data blocks within a \emph{stream ID}, each having a unique \emph{block ID}. Streams have associated \emph{metadata properties} as a set of name--value pairs, and is used in searching. Each block has a \emph{data payload} as a byte-array, and also \emph{metadata properties}.

\emph{Stream properties} include the stream ID, start and end time range of its blocks, sequence IDs of the blocks, and user-defined properties like sensor type, spatial location, etc. \emph{Block properties} are stream ID, block ID, sequence number, MD5 checksum, timestamp, and domain properties. Our store is optimized for append rather than update operations, with data and metadata often (but not always) \emph{immutable}.

While this model resembles other block and object stores like HDFS, Ceph and Azure Blobs, we additionally allow users to \emph{search over the block and stream metadata} to discover block IDs to access. This is useful when the IoT clients micro-batch sensor streams and create blocks with different temporal event ranges, and consumers wish to access blocks containing a particular time segment; or when different variables from the same sensor is placed in different blocks of a stream and users wish to access blocks holding specific variables. If need be, streams can be treated as directories and blocks as files within them to even offer a \emph{distributed file-system view}.

Given this, \elf supports the following service API:
\begin{itemize}
\item \texttt{CreateStream(sid, smeta[], r)} This creates a logical stream with ID \texttt{sid}, with \texttt{r} as the stream's reliability (i.e., reliability required for its blocks), and registers its metadata with the local (owner) fog, with an initial version number, and indexes it for searching. Metadata properties may be static or dynamic.
		
\item \texttt{Open|ReopenStream(sid)} This is optional\-ly used before \texttt{Put} to acquire an exclusive write lock to the stream for this client. Its response is the lease duration. \texttt{Reopen} renews the lease before it expires.
\item \texttt{PutBlock(sid, bid, bmeta[], data, lease)} Put adds a single new block \texttt{bid} to the end of the stream \texttt{sid}, with the given data payload and the stream's reliability, and registers its static block metadata for searching. If \texttt{lease} is passed from \texttt{Open} or \texttt{Renew}, it supports concurrent puts. Else, it behaves as an optimistic, lock-free protocol. 
\item \texttt{UpdateBlock(sid, bid, data, lease)} This updates the data contents for all replicas of an existing block, but is otherwise similar to put.
\item \texttt{UpdateStreamMeta(smeta[], v)} This allows the dynamic metadata properties for a stream to be updated, where \texttt{smeta} has the updated properties and \texttt{v} the version number of the old metadata being updated. 

\item \texttt{FindStream(squery)} This searches for streams that match a given set of static stream properties provided in the \texttt{squery}, and returns their IDs.
\item \texttt{FindBlock(bquery)} This searches for blocks that match a given set of static properties provided in the \texttt{bquery}, and returns their stream and block IDs.

\item \texttt{GetStreamMeta(sid, latest)} This fetches the cached metadata for the stream \texttt{sid} and their version. The \texttt{latest} flag forces the most recent version of the metadata to be fetched.
\item \texttt{GetBlock(sid, bid)} This downloads the data and metadata for the given stream and block ID.

\end{itemize}
Every fog runs a service that exposes these APIs, and clients can initiate an operation on any fog. These can be enhanced in future by APIs like \texttt{InsertBlock}, \texttt{GetBlockRange}, \texttt{GetBlockMeta}, \texttt{DeleteBlock}, \texttt{DropStream}, etc.

\subsection{Device Management}		
	
\subsubsection{Super-peer Overlay}
\elf uses a \emph{P2P model} for device management and search. Fogs act as \emph{super-peers} and edges as \emph{peers} within them~\cite{yang2003}. Each edge peer attaches to a single fog super-peer, which serves as its \emph{parent} and manages \emph{search and access} to its data and storage. A fog and its edges form a \emph{fog partition}. This reflects practical IoT deployments where such a 2-level hierarchy is common~\cite{he2017}. E.g., there may be a fog within a university campus, and all edges in the campus LAN are part of this fog partition.

Typical P2P networks scale exponentially, but require a \emph{logarithmic number of hops} to locate information~\cite{stoica2002}. Each (super)peer maintains routing details to $h$ (super)peers,  
where 
$2^h$ is the number of items that can be stored in the network. These form an \emph{overlay network} that takes up to $h$ \emph{hops} to locate a peer containing an item ID. Since we expect the fogs to number within the thousands and without a lot of flux, we instead maintain the super-peer overlay as a recursive 2-level tree. Each fog maintains a list of $b$ \emph{buddy} fogs at the first level (which form a \emph{buddy pool}), and a list of $n=(\frac{p}{b+1} - 1)$ \emph{neighbor fogs} at the second level, where $p$ is the total number of fog devices. Buddy pools are mutually exclusive, as are the neighbors of buddies in each pool. This limits our searches to $2$ hops -- first to a buddy and then to its neighbor~\footnote{This model can be easily extended to a classic super-peer overlay that scales to millions of fogs but with $h$ hops, or to support $b$-level redundancy for fog failures by having edges use all $b+1$ buddies as parent fogs~\cite{yang2003}.}.
Edges know which parent fog to join, and since our fogs do not come and go often, existing P2P discovery mechanisms or even simpler techniques can be used for constructing this overlay network.

\begin{figure}[t]
\vspace{-0.1in}
\centering

\subfloat[Fogs \& edges in a 2-level \emph{super-peer overlay network}]{
    \includegraphics[width=0.46\columnwidth]{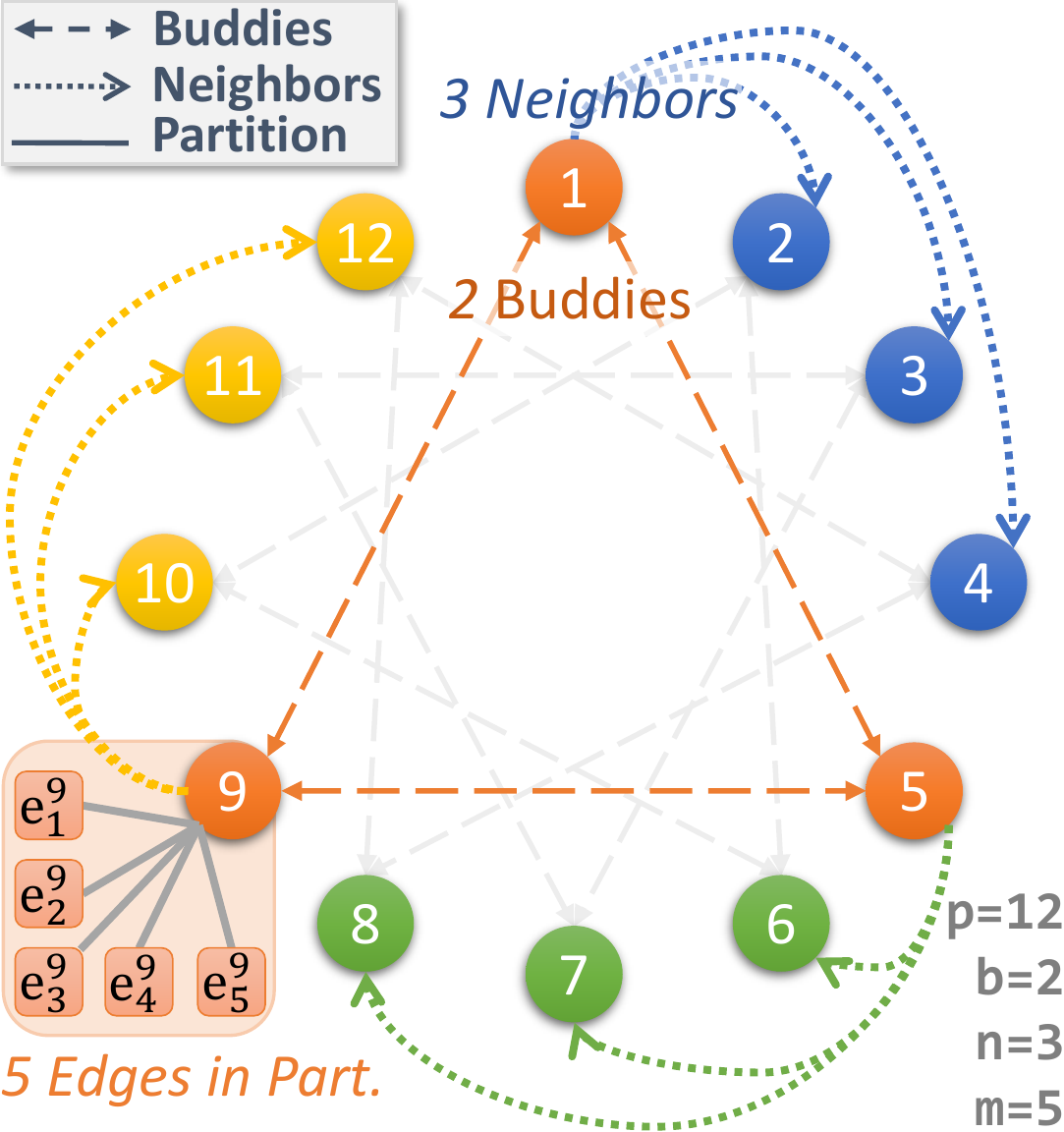}
    \label{fig:overlay}
  }~~
\subfloat[Update and search of \emph{federated index} at \emph{fog 1}]{
     \includegraphics[width=0.44\columnwidth]{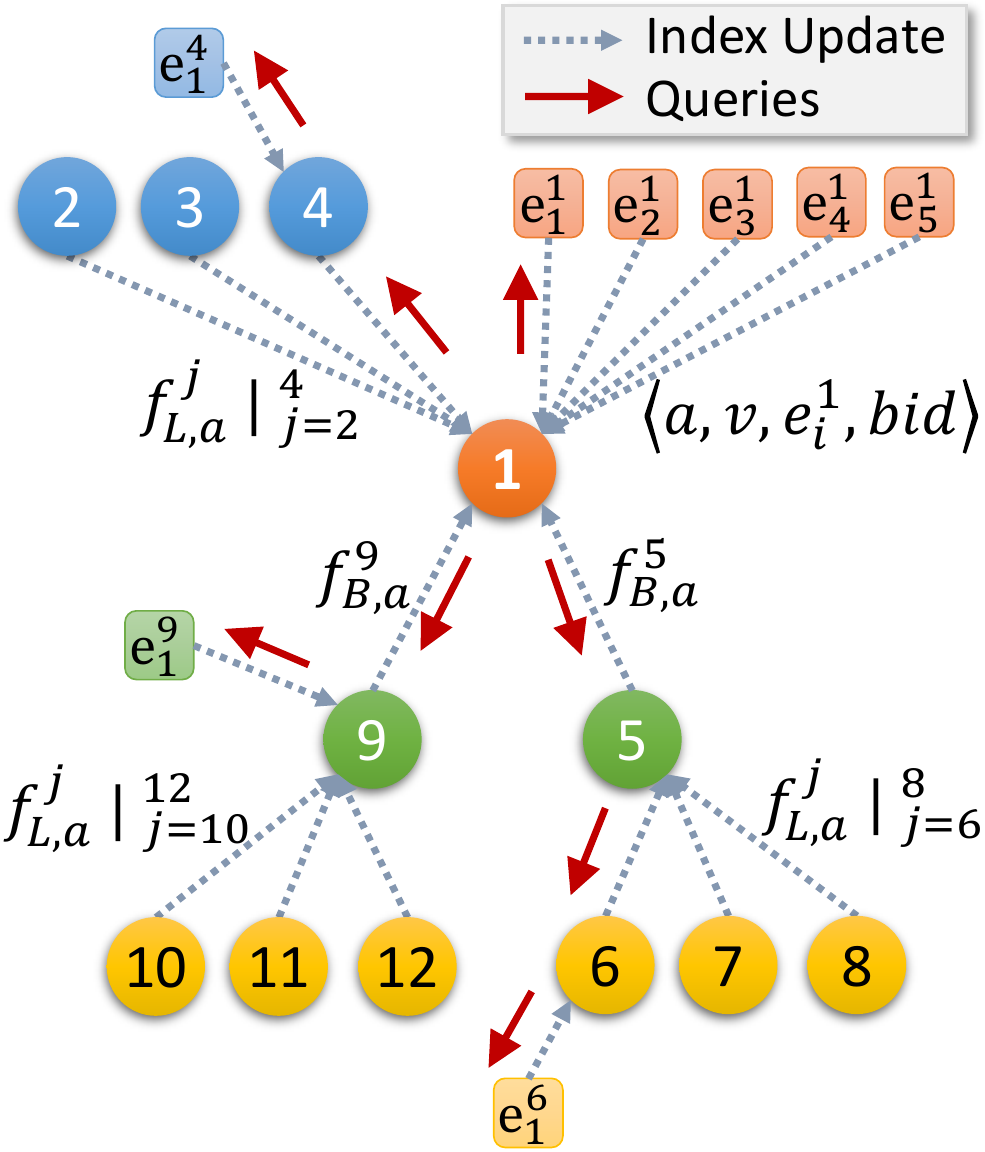}
     \label{fig:bloom}
  }

\caption{Overlay Network and Federated Index}

\vspace{-0.15in}
\end{figure}

Fig.~\ref{fig:overlay} shows $p=12$ fog super-peers in an overlay, each with $b=2$ buddies and the other fogs being partitioned across these buddies to give $n=3$ neighbors each. For brevity, neighbors for only one buddy pool and edges for only one fog partition are shown. E.g., \emph{fog 9} maintains details on its buddies \emph{1} and \emph{5}, neighbors \emph{10, 11} and \emph{12}, and edges, $e_1^9$--$e_5^9$.

\subsubsection{Health Monitoring and Statistics}
Light-weight \emph{heartbeat events} that are a few bytes long and sent often ($\approx 10$--$100~secs$) are used to monitor the devices. We also piggy-back tens of bytes of \emph{metadata and statistics} in these heartbeats.
This \emph{monitoring plane} enables fail-fast detection of device failures, and federated statistics to be maintained (Fig.~\ref{fig:arch}).

Each edge in a fog partition sends heartbeats to its parent fog when it is online, say every $30~secs$. The arrival or loss of an edge is detected using this. Multiple heartbeat misses indicate a loss, and will trigger \emph{re-replication} of blocks on the missing edge, while an edge arrival will make its storage available.
This obviates the need for a ``graceful'' entry or exit of edges. Fogs in a buddy pool send heartbeats to each other. Besides detecting the loss of a buddy and recovering its state (in future), this passes aggregate statistics from each buddy about its neighbors to others in the pool. Likewise, neighbors of a fog send it heartbeats and statistics periodically. Such heartbeats between buddies, and between neighbors and a fog, can help maintain the overlay network as fogs come and go.

\subsection{Data Discovery using Federated Indexes}

Typical P2P DHTs use consistent hashing over their IDs to locate the peer hosting the content. But we provide a unique feature to locate streams and blocks using their \emph{static metadata}, and not just ID. We maintain a \emph{federated index}, updated using the heartbeat events, to enable this (Fig.~\ref{fig:arch}). First, each fog maintains a \emph{partition index} of the metadata for blocks present in its edges and streams registered with it. This index is updated when a stream is created on the local fog that becomes its owner, or when a block replica is placed on it as part of an \texttt{PutBlock} call or a re-replication. 

Each edge $e_j^i$ sends a $\langle a,v,e_j^i,bid \rangle$ tuple to its parent \emph{fog i}, when a block $bid$ with property name $a$ and value $v$ is put on it\footnote{The block and stream IDs themselves are a property name. We use a similar approach for stream metadata, but omit its discussion for brevity.}. The fog maintains the index $\mathcal{I}_a:v \rightarrow (e_j^i, bid)$, that locates edges and block IDs in its partition that match a name--value pair. This update tuple is shown in Fig.~\ref{fig:bloom} for \emph{fog 1} from its edges, and allows the fog to answer \emph{0-hop} queries -- \texttt{FindBlock} queries over these property name(s) can be answered locally to return the matching block IDs and edges.

We also maintain a \emph{hierarchical Bloom filter} from neighbors, buddies and their neighbors to identify fog partitions that potentially host block(s) matching a given key--value pair, within \emph{2 hops} of the fog initiating the search request. Specifically, each \emph{fog i} applies its edge metadata updates to a \emph{local Bloom filter} for each property name, given as $f^i_{L,a} = \bigvee_k (\mathcal{H}(v_k))$, where $\mathcal{H}$ is a fixed bit-width multi-level hash function, $v_k$ are the set of distinct values for the property name \emph{a} for blocks present in this partition, and the Bloom filter is formed by a bitwise \texttt{OR} over all the hashes~\cite{broder2004}. We test if a value $v'$ is \emph{probably present} in the filter by checking if the bitwise \texttt{OR} of the filter with a hash of the value is non-zero, i.e., $(f^i_{L,a} \lor \mathcal{H}(v')) \neq 0$. 

Bloom filters can have \emph{false positives}, whose frequency is determined by the number of unique values inserted, the number of bits in the hash, and the quality of the hash~\cite{broder2004}. But it has \emph{constant-time} insertion and lookups, and \emph{compact storage}. In our experiments, we use a $160~bit$ SHA1 hash per property name.

Also, Bloom filters do not support deletions, and hence used to only index \emph{static properties} and not dynamic ones. This can be relaxed in future using \emph{Cuckoo Filters}~\cite{fan2014cuckoo}. 

When the local Bloom filter is updated, a fog sends it to other fogs it is a neighbor of, as part of the heartbeats. Each \emph{fog i} maintains list of $n$ \emph{neighbor Bloom filters} for a property name \emph{a}, one per neighbor \emph{fog j}, given as $\mathbb{F}^i_{N,a} = \{ \langle j, f^j_{L,a} \rangle \}$. This lets a fog check if any neighbor possibly contains blocks matching a given name--value query, and if so, forward the \texttt{FindBlock} query to the neighbor for an exact match using its local index $\mathcal{I}_a$. Fig.~\ref{fig:bloom} shows neighbors \emph{fogs 2, 3, 4} sending their updates to \emph{fog 1}, and responding to \emph{1-hop} queries.

Lastly, each fog encodes its local Bloom filters and its neighbor's Bloom filters into a recursive Bloom filter~\cite{ledlie2002}, 
and sends it to its buddies. For a \emph{fog j} with neighbors \emph{fog k}, this \emph{buddy Bloom filter} is constructed as $f^j_{B,a} = \bigvee_{k=1}^{n}(f^k_{L,a}) \lor f^j_{L,a}$. Each fog maintains $b$ buddy Bloom filters, $\mathbb{F}^i_{B,a} = \{ \langle j, f^j_{B,a} \rangle \}$, which allows it to test if its buddies or their neighbors possibly match a given query. E.g., in Fig.~\ref{fig:bloom}, buddy \emph{fog 9} constructs a buddy Bloom filter from its neighbor Bloom filters, \emph{fogs 10, 11, 12}, and its local Bloom filter, and passes it to \emph{fog 1}. This uses it for \emph{1-hop} (forward request to buddy) or \emph{2-hop} (forward to buddy's neighbors) queries.

Since client requests are routed through a fog, each fog maintains a \emph{cache of metadata} retrieved from others as part of various operations. This allows fast responses to other clients from the local fog's cache rather than the parent fog, but can return stale dynamic properties. Clients can pass a flag to force the latest metadata to be fetched. We do not cache data blocks to reduce the storage overhead, though it is a simple extension.

\subsection{Reliable Data Management and Access}

Each edge $e_i$ has a pre-defined \emph{device reliability} $r_i$, which can be part of the device specification or inferred from field experience. We also assume that blocks hosted on them are permanently lost when they disconnect from their parent fog.

\elf uses \emph{differential replication} to ensure that a block of \emph{size} $\bar{s}$ that it stores meets its \emph{block reliability} $\bar{r}$, by placing replicas on $q$ edges having \emph{available storage capacity} $s_i$ and \emph{reliabilities} $r_i$, such that $\bar{s} \leq s_i$ and $(1 - \bar{r}) \geq \prod_{i=1}^{q} (1 - r_i)$. So the replication count $q$ depends on both the reliability required for the block, and the reliabilities of the edges used. When a fog receives a request to put a block with its stream's reliability, it determines the replication factor $q$ and the exact edges to put these replicas on. E.g., a reliability of $\bar{r} = 0.999$ (i.e., $99.9\%$) can be achieved for a block by replicating it on $q=3$ edges with reliabilities, $r_i=\{ 0.80,0.91,0.95 \}$ such that $(1-0.999) \geq (1-0.80)\times(1-0.91)\times(1-0.95)$, 

or on $q=2$ edges having $r_i=\{0.95,0.99\}$.

The key challenge is that with $1000's$ of edge devices, it is not possible for each fog to maintain the current capacity and reliability of every edge device to make this decision. Instead, just as we used federated indexes to locate blocks, we similarly propagate and maintain \emph{approximate statistics} about the storage and reliability of edges in various fog partitions within the overlay network to help make this decision.

\subsubsection{Approximate Statistics} 

\begin{figure}[t]
\centering
\includegraphics[width=1\columnwidth]{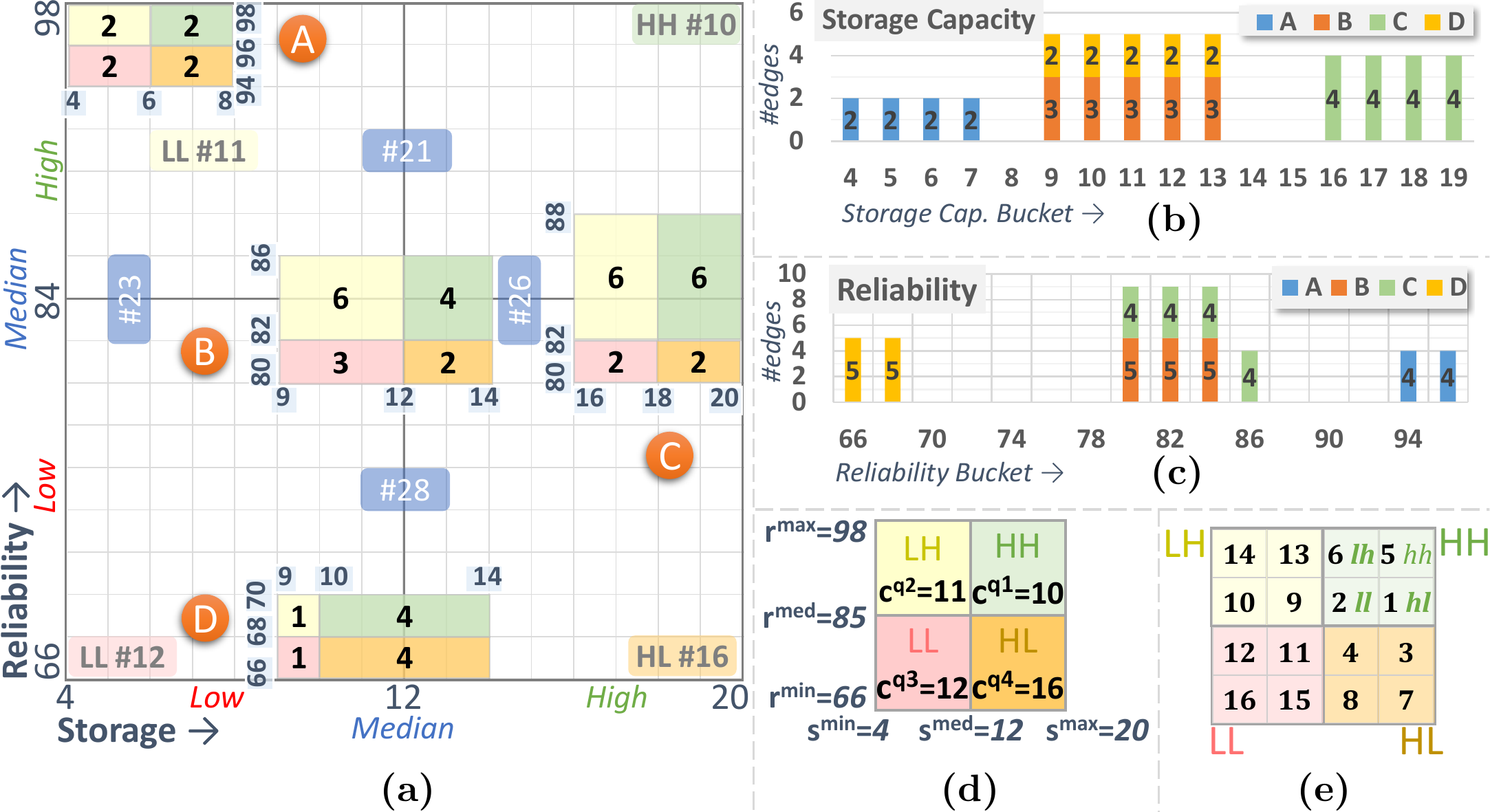}
\caption{Global Matrix Estimation for Storage and Reliability.
}
    \label{fig:stats}
\vspace{-0.15in}
\end{figure}

Each edge $e_i$ reports its reliability and available storage capacity $\langle r_i, s_i \rangle$ to its parent fog, periodically as part of its heartbeat. Each \emph{fog i} then determines the \emph{\underline{min}imum, \underline{max}imum and \underline{med}ian reliabilities and storage capacities} for all its edges, $\langle r_i^{min}, r_i^{med},  r_i^{max} \rangle$ and $\langle s_i^{min}, s_i^{med}, s_i^{max} \rangle$, along with the \emph{count} of edges that fall within each quadrant of this 2D space, $\langle c_i^{q1}, c_i^{q2}, c_i^{q3}, c_i^{q4}\rangle$, as illustrated in Fig.~\ref{fig:stats}(d). Here, we have $c_i^{q1}$ edges with reliability between $[r_i^{med},r_i^{max})$ and capacity between $[s_i^{med},s_i^{max})$; $c_i^{q2}$ edges with $[r_i^{med},r_i^{max})$ and $[s_i^{min},s_i^{med})$; and so on for the other 2 quadrants.

These edge counts correspond to the combinations of high/low capacity and high/low reliability, \texttt{HH, HL, LL, HL}. We will also have $c_i^{q1} + c_i^{q2} \approx c_i^{q3} + c_i^{q4}$, and $c_i^{q1} + c_i^{q4} \approx c_i^{q2} + c_i^{q3}$, depending on rounding errors.

These $10$-tuple values are then sent to the fogs we are a neighbor of, as part of heartbeats. Similarly, buddies exchange their neighbors' and their own tuples with other buddies. Using these $10$-tuples acquired from all fogs, each fog independently and consistently constructs a \emph{global distribution matrix}, as follows. We first find the \emph{global min and max} storage range among all the fogs, $s^{min} = \min_i{(s_i^{min})}$ and $s^{max} = \max_i{(s_i^{max})}$, and likewise the reliability range, $r^{min}$ and $r^{max}$. We divide each range $[s^{min},s^{max})$ and $[r^{min},r^{max})$ into $k$ equiwidth buckets, and for each \emph{fog i}, proportionally distribute its $(c_i^{q2} + c_i^{q3})$ count among the storage buckets that overlap with $[s_i^{min},s_i^{med})$, and its $(c_i^{q1} + c_i^{q4})$ count among buckets that overlap with $[s_i^{med},s_i^{max})$; and similarly, distribute counts $(c_i^{q3} + c_i^{q4})$ and $(c_i^{q1} + c_i^{q2})$ proportionally to reliability buckets that overlap with the reliability sub-ranges for the fog. We sum these bucket values across all fogs, and calculate the \emph{global median} storage and reliability, $s^{med}$ and $r^{med}$. This gives us the bounds of the global quadrants.

For the $10$-tuples for the $4$ fogs, \emph{A, B, C} and \emph{D} shown in Fig.~\ref{fig:stats}(a), their contributions to the storage and reliability buckets are shown in (b) and (c), using $k=16$ buckets. These help decide the global bounds in (d).

E.g., \emph{fog B} contributes it $c_B^{q2} + c_B^{q3} = 9$~edges proportionally to the $3$ storage buckets that fall between $[s_B^{min},s_B^{med})=(9,12]$, and $c_B^{q1} + c_B^{q4} = 6$~edges to the $2$ storage buckets that between $[s_B^{med},s_B^{max})=(12,14]$.
From these plots, we find the new global medians, $r^{med}=85$ and $s^{med}=12$.

Now, for each \emph{fog i}, we consider the \emph{area overlap} of each if its local quadrants with each of the global quadrants, and proportionally include the fog's edge count from that local to the global quadrant. E.g., in Fig.~\ref{fig:stats}(a), \emph{fog C} contributes all its edge counts in quadrants $c_C^{q3}=2$ and $c_C^{q4}=2$ to the global $c^{q4}$ which fully contains them, while the $6+6=12$ edges in its $q1$ and $q2$ local quadrants, which overlap with both the global quadrants $q4$ and $q1$, are shared proportionally in a ratio of $1$:$3$ between them. This gives the global count of edges in each of these four storage and reliability quadrants, \texttt{HH, HL, LL, HL}. Given this, a fog is mapped to the quadrant where its median-center falls. E.g., \emph{fog A} falls in \texttt{LL} and \emph{C} in \texttt{HL}.

\subsubsection{Replica Placement for Put} 

We use this information maintained independently but consistently on each fog to handle the \texttt{PutBlock} operation, invoked by a client on any fog. The fog receiving a put request for a block of size $\bar{s}$ queries the stream $sid$ to get its reliability, $\bar{r}$. It then selects a series of $q$ fog partitions, and chooses an edge within each for placing a replica such that we (1) balance the use of fogs with both high and low reliability edges to ensure that \emph{a sustainable mix of edges remain}, (2) give preference to fogs that have a higher available storage to ensure \emph{effective use of capacity}, (3) select different fogs for each replica to enhance \emph{partition-tolerance} and \emph{locality} with diverse clients, (4) bound the \emph{replication factor} to a minimum and maximum value set by the user, and (5) meet the block's \emph{reliability requirement}.

We select fog partitions from different quadrants in the global matrix in a particular sequence to meet the above goals. Specifically, we alternate between \texttt{HH} and \texttt{HL} quadrants to prioritize high-capacity fogs. Within the global quadrant, we pick a random fog and test if it has a non-zero edge count in a complementary reliability quadrant. E.g., for a fog that maps to the \texttt{HH} quadrant of the global matrix, we check for edges in its \texttt{HL} or \texttt{LL} local quadrants, and for a fog in the \texttt{HL} global quadrant, we test for edges in its \texttt{HH} or \texttt{LH} local quadrants. If the fogs have zero edges in these quadrants, we expand to the other two local quadrants as well. 

The sequence order of global and local quadrants that are tested is given in Fig.~\ref{fig:stats}(e), and a variant of a \emph{Z-order curve}. Intuitively, \emph{this picks edges close to the median global reliability and with high capacity}. The reliability is initially met by median edges. As their capacity is exhausted, the edges with more extreme (low or high) reliability move closer to the median and will be chosen. Later, this helps us find pairs of edges with low and high reliability that together give a reliability similar to the initial two median reliability edges. As an optimization, we always try and place the first replica locally, if the writing client is on an edge. We also pick edges in different fog partitions unless there is no available capacity. 

A \emph{fog i} that is chosen will provide a minimum reliability of $r_i^{min}$ if the edge is in the \texttt{HL} or \texttt{LL} local quadrant, or $r_i^{med}$ if in \texttt{HH} or \texttt{LH}. This is a conservative estimate since the actual edge selected within the fog may have a reliability as high as $r_i^{med}$ or $r_i^{max}$, respectively. We pick as many fogs as needed to meet the block's reliability or the minimum replication count.

The fogs chosen in this manner are sent to the client, which then directly contacts each fog \emph{concurrently} to place a replica of the block. Each fog selects an \emph{edge with the least reliability in the specified local quadrant}, and puts the block on it. In case the global matrix is stale and the fog cannot find a suitable edge, this fog can use its own global matrix to find an alternative fog with a similar non-empty global and local quadrant. Since the edge may be on a private network, the \emph{data moves} from the client to the parent fog hosting a replica, and from it to the edge. If the client is an edge, it will also pass through its own parent fog first, but not otherwise, to avoid the extra hop. The fog also \emph{registers the block metadata} with itself, propagates to the federated indexes as described before, and updates the \emph{stream metadata} at the owner fog with the block ID, MD5 checksum, and block count.

\subsubsection{Getting a Block} 

Getting a block involves finding the fogs containing the block replicas using its ID from the local fog. This first returns the \emph{local fog} or the possible \emph{neighbor fogs} that may contain it, based on a local index or Bloom filter lookup. The client contacts the local fog if present in the response, and this will have the replica. Else, the client contacts each neighbor fogs, which checks its local index, and if present, returns the block from the edge to the client. 

If none of the local or neighbor fogs hold a copy, or in the rare case these were all false positives, we recheck with the local fog and force it to search its buddy Bloom filters. It forwards the find request to matching buddies to check their local index and neighbor Bloom filters, in \emph{$2$--$3$ hops}. This will return the global list of fogs that may contain the replica, and the client contacts each to get the first available replica.

\subsubsection{Re-replication for Recovery}
A parent fog detects an edge failure due to missing heartbeats. This triggers a recovery of all block replicas present on the edge to ensure each block's \emph{reliability requirement} is still met. For this, the fog uses the same edge selection approach as above, except that it tries to find a single fog that has an edge with a reliability similar to the edge that failed. The parent fog then gets an existing block replica from a surviving edge, and puts it on the newly selected fog and an edge within it. This selection of alternative devices and re-replication onto them is done concurrently for lost blocks on the failed edge. While we currently assume that the reliability for an edge does not change over time, in future, this same technique can be extended to expand or contract the number of replicas to adapt to dynamism in the reliability.

\subsection{Consistent Concurrency and Updates}

\subsubsection{Concurrent Puts and Updates with Leasing}
	
The default \texttt{PutBlock} operation is optimistic, and assumes that just one client is writing to the stream. 
With concurrent clients adding blocks, the order in which the blocks are appended to the stream depends on the order in which the stream metadata at the owner fog is updated with the new block IDs. Here, we will need a user-defined sequence number in the block metadata for \emph{partial ordering} of blocks written by one client.

However, for global ordering of blocks with concurrent clients, we offer a \emph{soft-lease mechanism}.
Here, the client first calls \texttt{OpenStream} to try and acquire a lease for a certain duration. This request is forwarded to the owner fog for the stream, which logs and returns a successful lease for the requested (or a pre-defined) duration, if no other client holds an active lease on this stream. The response has the \emph{duration} and a \emph{session key}, which is a unique random nonce used for auditing. \texttt{PutBlock} then passes the client ID, lease duration and session key to the fogs where the replicas will be placed. These fogs sanity-check if the lease duration is valid, and log the client ID and session key for this operation, before writing the block replica to their edge. The client also adds the new block IDs to the stream metadata.

This \emph{soft-lease model} is light-weight, but does not \emph{enforce} locking of the stream. 
It is up to the clients to ensure that they have acquired a valid lease before they call puts in parallel to avoid inconsistent ordering. But, the logs maintained at the fogs allow us to later verify the validity of the operations.

The lease on a stream can be used by the client across multiple \texttt{\\ Put|UpdateBlock} operations. This lets it write a series of blocks to the stream \emph{with guaranteed contiguous order}. If the lease is going to \emph{expire} before an operation, the client \texttt{Renew}s it with the fog, which returns an extended lease duration if it has not expired. If the lease has expired \emph{and no other client has acquired the lease since then}, the fog goes ahead and extends the lease. This reduces leasing overhead dues to time-skews, without affecting consistency. If an \texttt{OpenStream} fails due to another client having the lease, the client can poll and retry acquisition. There is no explicit close stream operation, and the lease is released on expiry.

\texttt{UpdateBlock} is similar to \texttt{PutBlock}, but replaces the selection of replicas using the global matrix, with finding the fogs holding all the current replicas for the block, similar to \texttt{GetBlock}. Once located, the client sends the updated block data to each replica, and also updates the stream metadata with the new MD5 checksum for the block.

\subsubsection{Stream Metadata Updates}

When a stream is created, it is registered with an \emph{owner fog} that holds it metadata. These properties may be static or dynamic. While static properties are indexed and searchable, the values of dynamic properties can be updated but not searched on.

Leasing is useful when multiple operations are done with a single lease to amortize its cost. But metadata updates are single operations. So we assign version numbers to dynamic metadata properties and employ a \emph{test and set} pattern to allow consistent and concurrent updates to them. 
This version is returned by \texttt{GetStreamMeta}. Cached versions of the stream metadata also maintain and return the version in their cache.

When updating the metadata for a stream, the client first does a \texttt{GetStreamMeta}, updates the values of the returned dynamic properties, and sends the new property values and the \emph{earlier version number} to the owner fog of the stream. The fog \emph{tests} if the current version matches the passed version, and if so, \emph{sets} the passed dynamic properties and increments the version. But, if the current version is greater than the one that is passed, then the client is trying to update a stale copy of the dynamic property. This may be due to using an older cached metadata on a different fog, or another client having updated the metadata with the owner fog since the last access by this client. Then the update call fails, and the client has to get the latest metadata and retry with the new version number.

There are also system-defined dynamic properties that are maintained as part of various APIs, such as the block count, list of block IDs, and their MD5 checksums, for a stream. These cannot be modified directly by the client, but the framework updates these internally using a similar pattern.

\section{Experiments}
\label{sec:results}

\elf is implemented in Java using the Apache Thrift cross-platform micro-services library. The fog service has the bulk of the logic, while the edge services are light-weight.

We conduct experiments to validate the performance, resilience and scalability of \elf. We use the \emph{VIoLET} container-based IoT virtual environment to define two deployments~\cite{violet}. In the first, \textbf{D20}, we have $4$ fog containers on a public network, with $4$ edges connected to each fog in a private network. This gives a total of $20$ devices running on $4$ Azure D32 VMs ($32$-core, $128$~GB RAM). The \textbf{D272} configuration has $16$ fogs, with $16$ edge containers each, for a total of $272$ devices on $1$ public and $16$ private networks. They run on $16$ Azure D32 VMs. All devices in each fog partition run on the same VM. The edge containers have CPU and memory resources that match a Raspberry Pi 3B ($4$-cores@$1.2$~GHz, $1$~GB RAM, $16$~GB disk space), while the fog containers map to a Jetson TX1 ($4$-cores@$1.9$~GHz, $4$~GB RAM), as defined in VIoLET.  Network links have a bandwidth of $90$~Mbps.
We use a Normal distribution for the edges' reliability, with $\mu=90\%, \sigma=3\%$ for D20, and $\mu=80\%, \sigma=5\%$ for D272.

\begin{figure*}[t]
\centering

\subfloat[\texttt{PutBlock} without leasing]{
       \includegraphics[width=0.65\columnwidth]{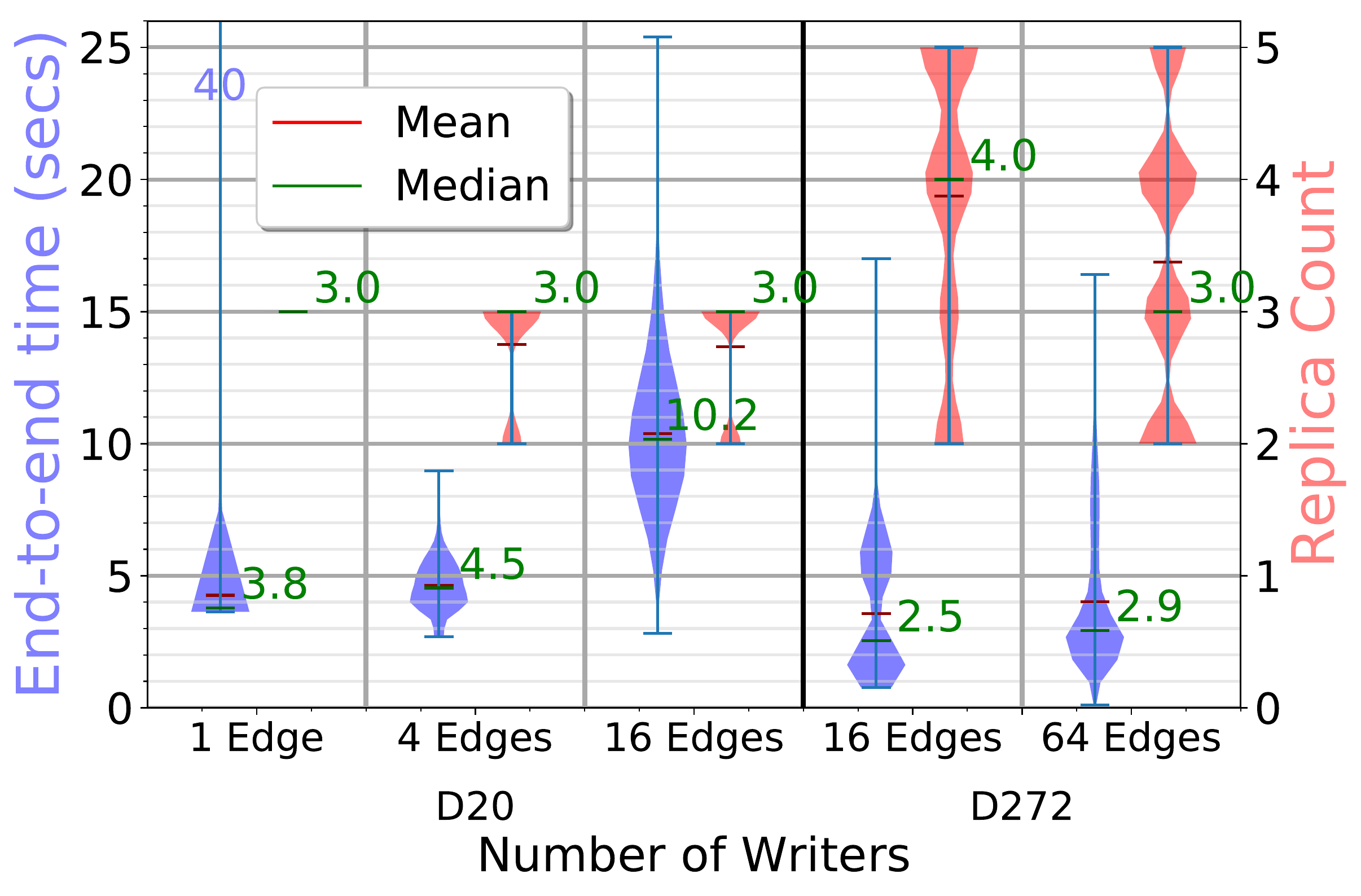}
    \label{fig:update_stream_metadata_contention}
\label{fig:write}
      \label{fig:d272:write}
      \label{fig:d20:write}
  } 
    
\subfloat[\texttt{PutBlock} on \emph{D20} with leasing \& concurrent writers]{
~\includegraphics[width=0.65\columnwidth]{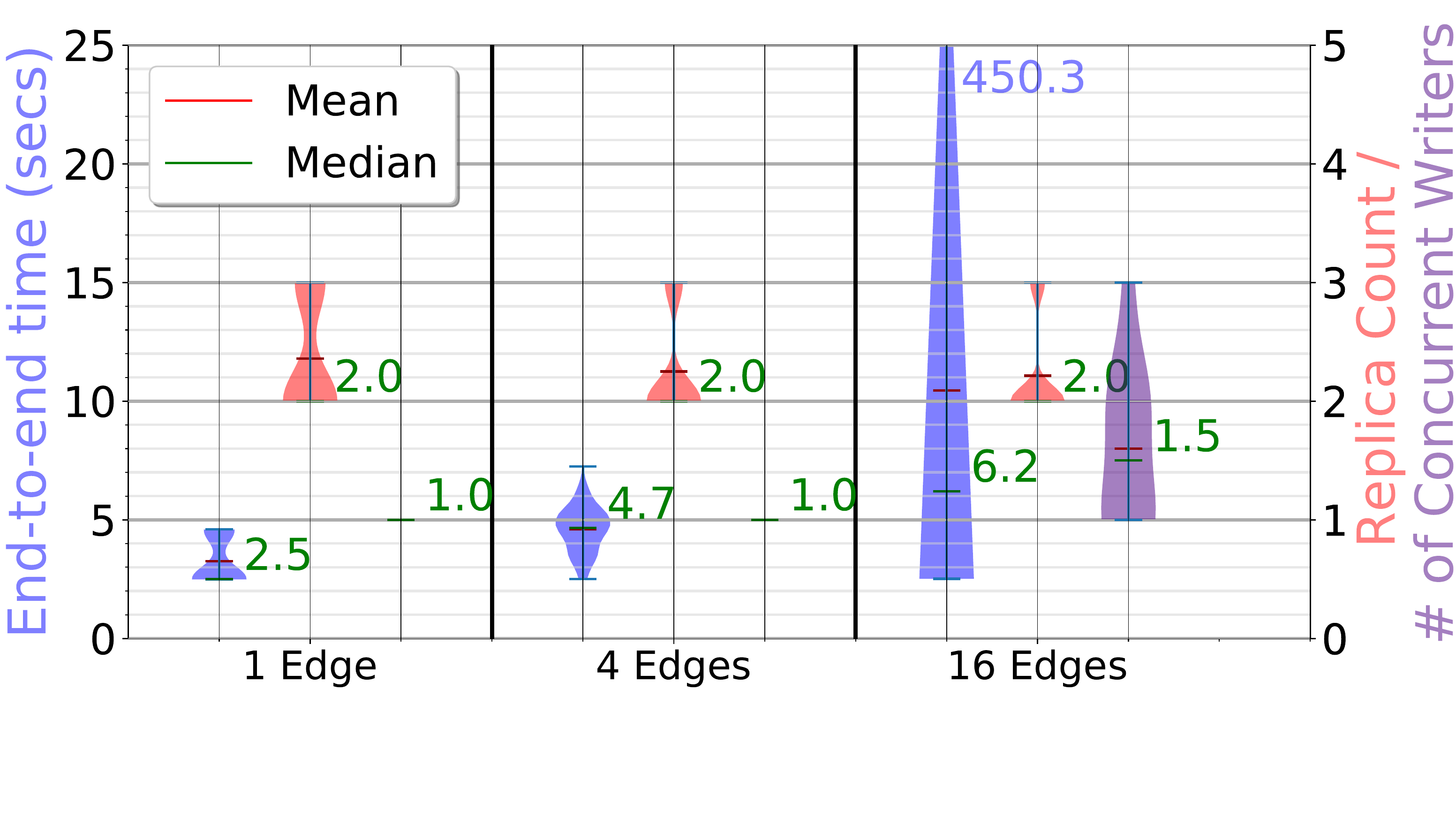}~
    \label{fig:block_appends_locking}
  }

\subfloat[\texttt{Find} and \texttt{GetBlock}]{
      \includegraphics[width=0.65\columnwidth]{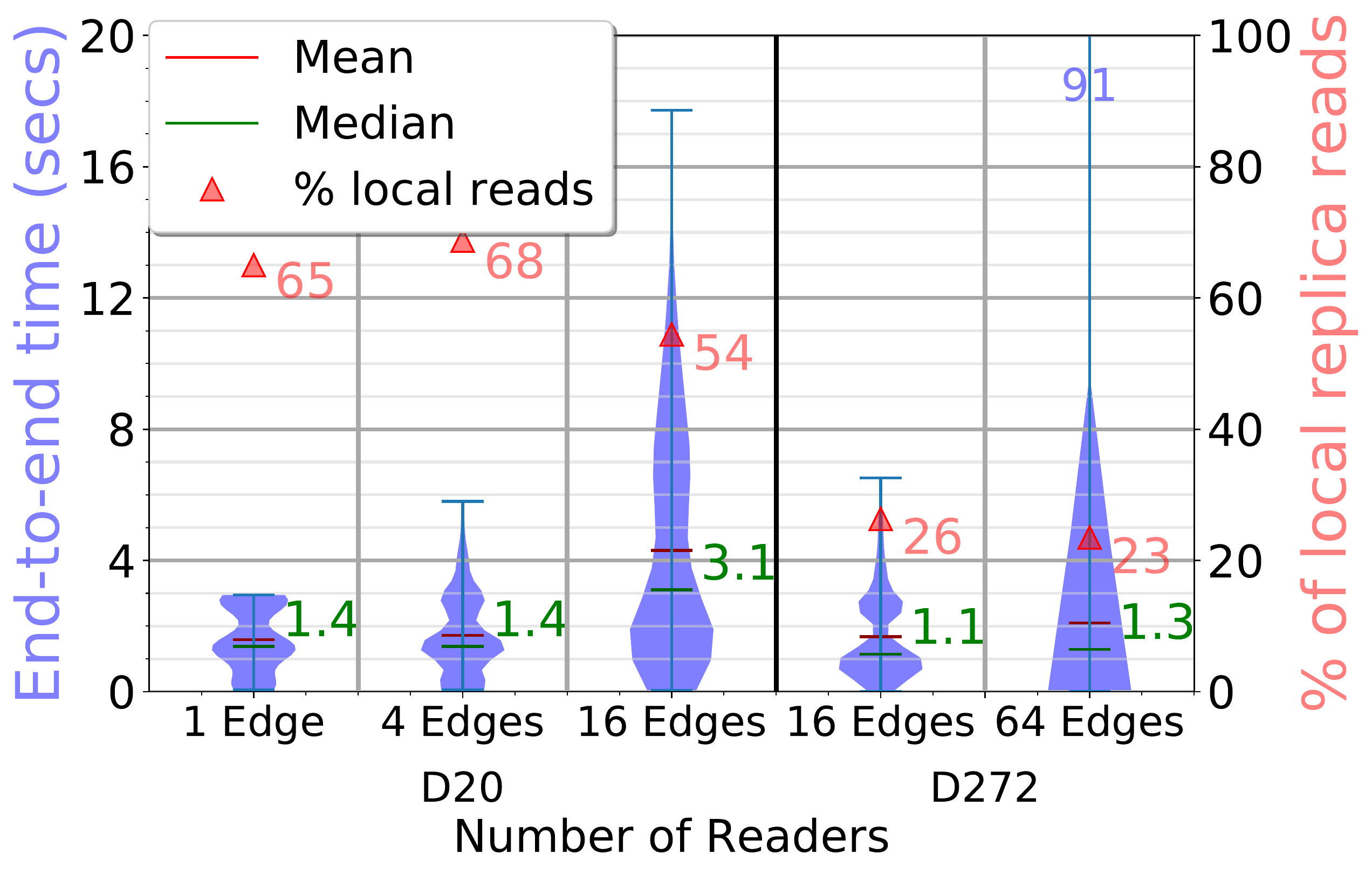}
    \label{fig:read}
      \label{fig:d20:read}
      \label{fig:d272:read}
  }
\caption{Performance of Put and Get block operations}

\end{figure*}

\subsection{Put Block Performance}
\subsubsection{Put performance without leasing} 

For the \emph{D20 setup}, we run experiments with $1, 4$ or $16$ edges concurrently calling the \texttt{PutBlock} API on their local fog parent with a blocks size of $10~MB$, in a loop for $100$ times. We set a reliability of $\bar{r}=99\%$ for all these streams, and a min and max replication factor of $2$ and $5$. For the \emph{D272} setup, we perform two experiments with $16$ and $64$ concurrent edge clients spread across the $16$ fogs. Each edge calls \emph{put} for $100$ iterations. They put blocks of size $1~MB$ or $10~MB$ and use reliabilities of $90.00\%, 99.00\%, 99.90\%$ or $99.99\%$, with uniform probability. This diversity reflects realistic scenarios. Leasing is not enabled, and edges put to distinct streams in their local fog; one replica will be placed in the local edge.

The \emph{end-to-end latency distribution} in seconds for the \emph{put} API calls is shown as blue violin plots on the left Y axis of Fig.~\ref{fig:d20:write}. For a single API call, this is the time to \emph{find} the fogs to place block replicas on, copy all replicas to the target edges concurrently, and register the block metadata. Each violin distribution has $\# edges \times 100$ data points.

For \emph{D20}, with \emph{1 edge writing}, each \emph{put} call takes a median of $3.8~secs$. Since each replica is $10~MB$ in size, the link speed is $90~Mbps$, and we need $3$ hops -- from client to parent fog, parent fog to target fog, and target fog to edge -- about $3~secs$ are spent just in data movement. 

Zooming in, the time to \emph{find} the replica placement is just $30~ms$ as the parent fog takes a local decision, and the time to update the metadata index is also $30~ms$; this is mostly the service call overhead. 

These times do not vary much as we increase to $4$ concurrent edges writing 
from $4$ different fog partitions, with their median time at $4.5~secs$. But with $16$ edges putting blocks  in parallel, all $4$ edges of every fog are active. Since they all route data through their parent fog to a remote fog, the data transfered out from the fog for edges in its partition is $4~\text{\em edges} \times 2~\text{\em remote replicas} \times 10MB$. Hence, its available bandwidth limits the performance, taking a median of $10.2~secs$.

So \elf's \emph{overheads are minimal} in all these cases, and we are only \emph{bandwidth bound}.

For \emph{D272}, each edge is randomly assigned to put blocks of either $1MB$ or $10MB$ in size, $100$ times. For \emph{16 edges}, there are $8$ edges each putting blocks of these two sizes, while 
for \emph{64 edges}, there are $25$ writing $10MB$ blocks and the rest $39$ writing $1 MB$ blocks. Fig.~\ref{fig:write} shows that the median latency with $16$ concurrent edges is about $2.5~secs$ and it only marginally increases to $2.9~secs$ for $64$ edges. The smaller time than D20 is due to the use of smaller block sizes and a smaller client load, compared to the total edge count.

\clearpage

If we limit our analysis to just the edges on \emph{D272 putting $10MB$ blocks} 
(plots omitted for brevity), we report that the median time for the $8$ (of $16$) edges writing $10MB$ blocks is $5.5~secs$ while for $25$ (of $64$) edges it is $6.8~secs$.
These are higher than D20 primarily due to the higher replication factor, which has grown from being $\approx 3$ to as high as $5$, as seen in the red violin on the right Y axis of Fig.~\ref{fig:d20:write}. This increases the data transfer time, both due to additional bandwidth and the compute cost of concurrent threads doing these operations. 

The higher replication factor and its wider distribution for D272, spanning the full range of $2$--$5$ copies allowed, are due to its lower and more variable edge reliability of $\mu=80\%, \sigma=5\%$. In contrast, D20's reliability of $\mu=90\%, \sigma=3\%$ results in a replication factor of $2$--$3$.
This clearly shows the \emph{differential replication} at work.

\subsubsection{Put performance with leasing}

We initialize the \emph{D20 setup} with $16 \times 100$ block writes without leasing. Then, we perform $25$ \emph{additional} block puts per client to a random stream, from $1$, $4$ and $16$ concurrent clients, \emph{with a lease acquired} on the stream for $100~secs$, and renewed a median of $2$ times.

Different edges may select the same stream to write to. 
Besides the end-to-end latency for these leased-puts, which now includes the lease acquisition and renewal time (left Y axis in Fig.~\ref{fig:block_appends_locking}), and the replica count (right Y axis), we also show the concurrent writers count for a stream (right Y axis).

With $1$ or $4$ edges doing puts, we see that the median latency is $2.5~secs$ and $4.65~secs$. These are comparable to the previous experiments without leasing for the same number of writers. This is due to the lower median replication factor of $2$ in these runs (compared to $3$ earlier). This is due to a higher overall reliability of the edge devices in these runs, despite sampling from the same edge reliability distribution. No two edges have selected the same stream to write to in these runs. This indicates that the edge reliabilities, replication count and bandwidth usage have a bigger impact on the end-to-end latency than the leasing overheads.

\begin{figure*}[t]
\centering
\subfloat[\texttt{Get} and \texttt{Update\-Stream\-Meta}]{
     \includegraphics[width=0.38\columnwidth]{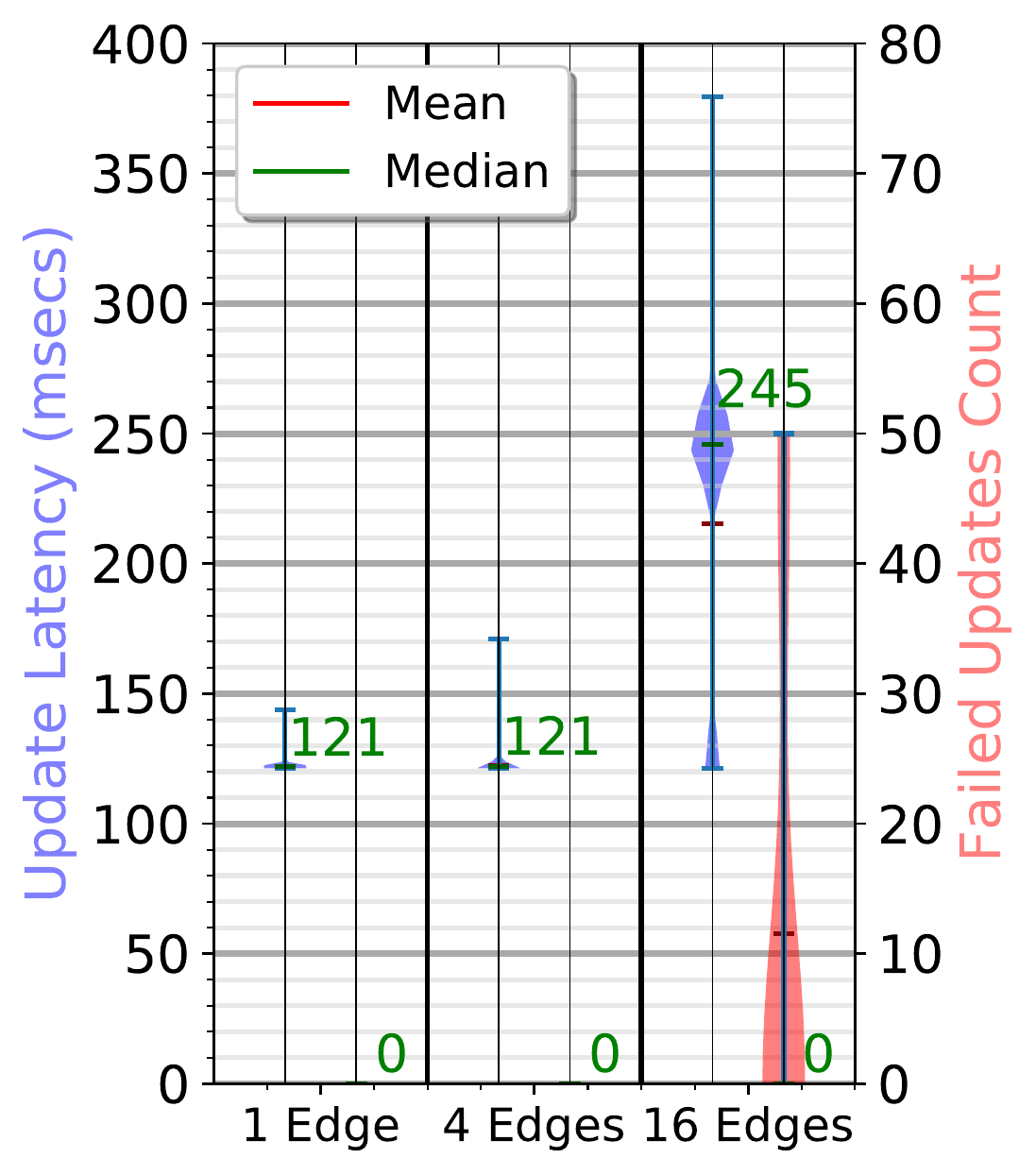}
    \label{fig:update_stream_metadata}
}~
\subfloat[D20 \& D272 recovery time and block count]{
     \includegraphics[width=0.55\columnwidth]{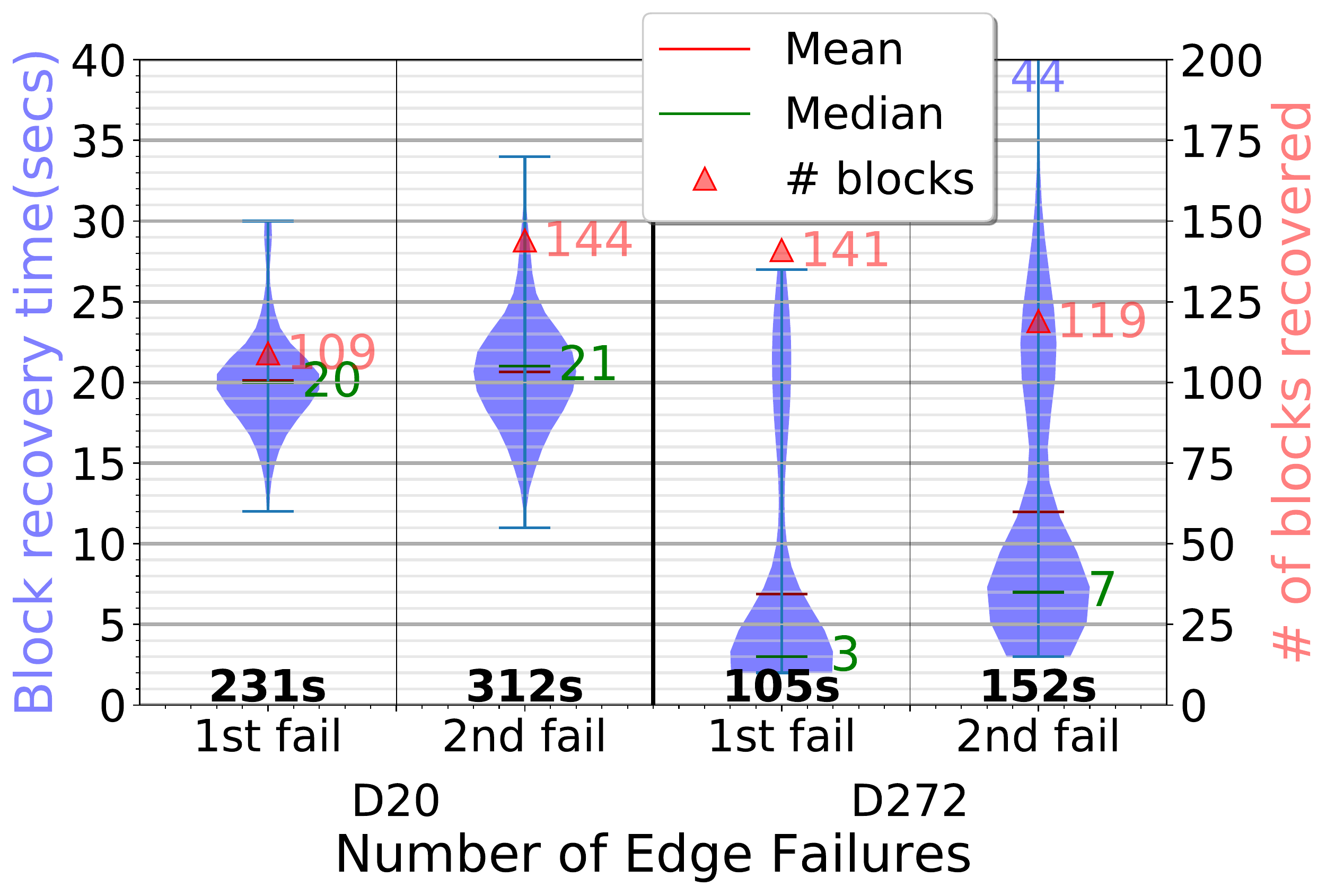}
      \label{fig:d20:recover}
}
\vspace{-0.1in}
\subfloat[\emph{Global Matrix} values over time, for \texttt{PutBlock}s (time $0$--$12$, circle) and two edge recoveries (times $15$ and $22$, cross) on D20]{
      \includegraphics[width=0.9\columnwidth,trim={.2cm .1cm .2cm .5cm},clip]{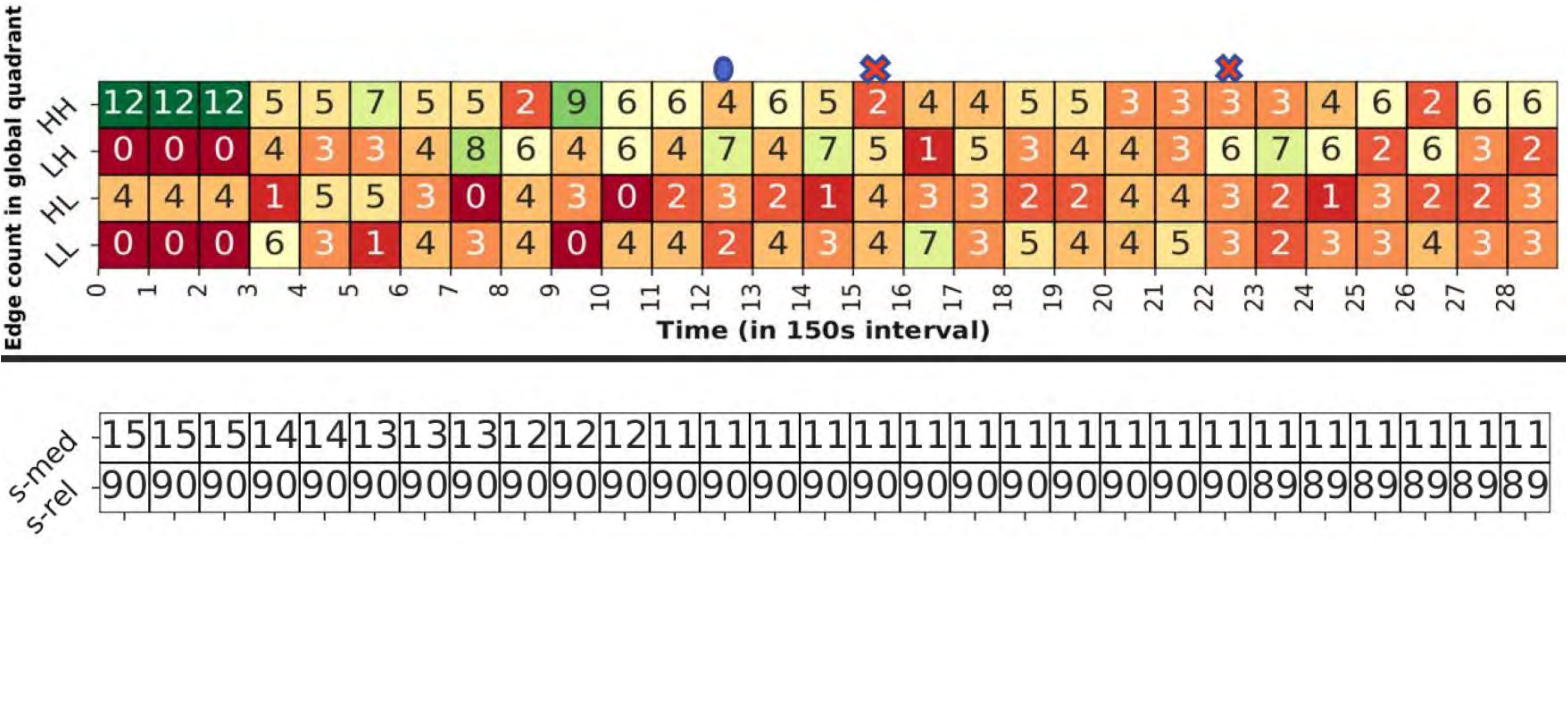}
       \label{fig:d20:heat}
}

\caption{Performance of stream metadata update, and block recovery after edge failures.}
    \label{fig:recover}
\end{figure*}

With $16$ clients, the median latency is lower than without leasing at $6.2~secs$ due to the smaller replication count. But the latency distribution is much wider, reaching $450~secs$. This is because multiple edges pick the same stream to write to, as seen in the right-most violin. We have 
$4$ streams selected by $2$ edges each to write to, and $1$ stream picked by $3$ edges. Hence, with concurrent writers and leasing, only one will write to the stream at a time while the others poll to acquire the lease. This lasts till all $25$ blocks are put by an existing edge with the lease.

The peak latency to write a block is for the stream with $3$ clients. The last edge to get the lease was waiting for $50$ blocks to be written by the previous two edges, that takes about $446~secs$. So the latency for this edge to put its first block is $450.3~secs$, while putting the rest of its $24$ blocks does not have additional leasing overheads.

\subsection{Find and Get Block Performance}

We do a similar set of concurrent \texttt{FindBlock} and \texttt{GetBlock} API calls from $1$, $4$ and $16$ edges for the D20 setup, and from $16$ and $64$ edges for D272. \elf has been loaded with $16 \times 100$ blocks (D20) or $64 \times 100$ blocks (D272) using the previous \emph{put} runs. Each edge \emph{finds} $100$ random block IDs from the ones inserted, followed by a \emph{get} of that block. 

The time to \emph{find and get} each block is shown in Fig.~\ref{fig:d20:read} (left Y axis), and a magenta triangle on the right Y axis indicates the percentage of times a \emph{replica from the local partition} is read. The \emph{find} API call is fast, taking about $220~ms$ with $1$ and $4$ edges for \emph{D20}, and about $440~ms$ with $16$ edges. In the latter case, each fog is servicing $4$ concurrent edge requests and hence marginally slower. 

Once the replicas for a block ID are identified, we \emph{get} one of the replicas -- preferring a replica in the local fog partition, if present. For \emph{D20}, we see that the \emph{get} latencies have a bimodal distribution. There are peaks at $1.4~secs$ and $2.6~secs$ for $1$ and $4$ edges, and at $3.1~secs$ and $7.5~secs$ for $16$ edges. This is due to the mix of local and remote replicas that an edge accesses. Edges are able to \emph{get} a local replica copy $55$--$70\%$ of the time, resulting in the lower latency peak. This range is within the $\frac{1}{4}\times 1 + \frac{3}{4}\times \frac{1}{4}\times 2 = 62.5\%$ we expect -- since all edge clients put blocks uniformly, $\frac{1}{4}^{th}$ of all the blocks have their first replica locally; of the remaining $\frac{3}{4}^{th}$ of blocks, there is a $\frac{1}{4}$ chance on the $\approx 2$ non-local replicas to be on that fog. The second peak reflects the copying of a remote replica. Just like for the write, we are bandwidth bound as the concurrency increases, showing that \elf has low overheads.

The performance for \emph{D272} is equally fast, taking a median $1.1$--$1.3~secs$ with $16$ or $64$ edge readers. It benefits from $50$--$60\%$ of blocks being only $1~MB$ in size. However, this is despite only $\approx 23\%$ of blocks having a local replica out of the median 4 replicas per block. This too matches the expected local fraction of $\frac{1}{16}\times 1 + \frac{15}{16} \times \frac{1}{16} \times 3 = 23.8\%$. In fact, the small number of local copies means that the latency distribution is tighter. So \elf weakly scales for \emph{gets} too.

\subsection{Metadata Update Performance}

We conduct experiments on the \emph{D20} setup to measure the latency for stream metadata updates, using $1$, $4$ and $16$ concurrent edges as clients. Each client randomly picks one of the $100$ existing streams, and performs $100$ \texttt{GetStreamMeta} and \texttt{UpdateStreamMeta} operations alternately on it. It is possible for two clients to select the same stream to perform an update. Since we use version checking rather than leasing for metadata updates, it is likely that the version of a stream metadata being updated may have been updated by a concurrent client and hence fail. We report the latency for \emph{get and update metadata} (left Y axis) and the count of failed updates (right Y axis) in Fig.~\ref{fig:update_stream_metadata}; failed updates are not retried.

With just $1$ or $4$ clients, no two streams are randomly picked for update by the same client, and only local streams are chosen. So all updates are at the local fog, and complete successfully with a median latency of $121~ms$.

But with $16$ clients, $4$ streams are selected by a pair of clients to update concurrently. This causes $185$ of the total of $1600$ updates to fail due to staleness, as seen in the right Y axis. The update time also increases to a median of $245~ms$. This is primarily due to a majority of the metadata updates happening on a remote fog partition, unlike the $1$ and $4$ edge cases, and this causes an extra network hop in the VIoLET environment.

\subsection{Block Recovery Performance}

Lastly, we measure the responsiveness of \elf in recovering from edge failures, and ensuring that the blocks maintain their reliability levels. We load $16 \times 00$ and $64 \times 100$ blocks 
into the D20 and D272 setups, like before, and then kill one of the edges with the least reliability. We track the time taken by its parent fog to detect the loss, and start re-replicating the lost blocks on other edges. Once recovery is complete, we kill another low reliability edge. Fig.~\ref{fig:d20:recover} plots the \emph{time to re-replicate} each block on the left Y axis violin, the number of \emph{blocks recovered} on the right Y axis, and list the \emph{total recovery time} at the bottom, shown after the first and the second failures.

In all cases, $100\%$ of lost blocks are re-replicated. 

We see that the re-replication time per block is $\approx 21~secs$ for \emph{D20}, and $\approx 3$--$8~secs$ for \emph{D272}. These are comparable to the sum of the \emph{get} 
and \emph{put} 
times seen before, 
since we \emph{get} a surviving replica and \emph{put} it on a new edge. Also, recovery of blocks is done in parallel on the fog using $10$--$20$ threads. Hence, while $109$--$144$ blocks are recovered depending on the failing edge, 
the total recovery time is only $105$--$312~secs$. So the thread parallelism gives us a $10\times$ speedup.

We further examine how our \emph{global matrix} changes as blocks are populated in \elf, and when failures happen. Fig.~\ref{fig:d20:heat} shows a heatmap of the edge-counts in the $4$ global matrix quadrants (top $4$ rows) and the median storage and latency values (bottom $2$ rows), updated every $150~secs$ along the X axis, for D20. At time steps $0$--$12$, $4$ edges are concurrently writing $100$ blocks in a loop. Initially, the median available storage $s^{med} = 14~GB$, and all $16$ edges fall in the high capacity quadrants, \texttt{HH} or \texttt{HL}. As replicas are written to fogs in these quadrants and their edge capacities get used on a priority, the count shifts from \texttt{HH} and \texttt{HL}, to \texttt{LH} and \texttt{LL}, e.g., from step $2$ to $3$. Eventually, this disk usage causes the median capacity to change, say, from $14~GB$ to $13~GB$ after step $5$. This causes borderline fogs, earlier classified as low capacity, to move to the high capacity, and become prioritized for selection. So we keep selecting fogs that are in and around the median value.

After step $15$, there is an edge failure and the total edge count drops from $16$ to $15$. The ensuing re-replication causes the missing blocks to be copied to an existing edge. While only one replica is created, this is done by $10+$ concurrent threads. So the edge counts again shift from high to low capacities. When a second edge fails after step $22$, it even causes the median reliability to drop from $r^{med}=90\%$ to $89\%$.

\section{Conclusions}
\label{sec:conclusions}

In this paper, we have presented a novel distributed storage service for edge and fog resources that offers a transparent means for edge computing applications to access streams of data blocks persisted locally. This avoids the need to move IoT data to and from the cloud, other than for long-term archival. \elf leverages ideas from both P2P networks and Big Data storage like HDFS. It uses a federated index for $2$-hop searching of blocks, with hierarchical Bloom filters over static metadata properties for fast probabilistic searches at scale. It maintains approximate global statistics on storage and reliability distributions of edges on different fogs, which helps it select fogs and edges for differential replication. This guarantees tunable reliability of each block. Our experiments demonstrate the low overhead of \elf, with block read and write performance bound only by the network speed. Consistent and concurrent updates of blocks and metadata are also validated. It also performs automated and rapid block re-replication on edge failures, to maintain the required reliability.

As future work, we plan to include support for overlay creation, as available in existing P2P literature, and use buddy pools to handle unreliable fogs as well. 
We can also enforce the leases as locks, and support access control, auditing and non-repudiation mechanisms. 
Larger scale and comparative experiments, and concurrent-failure tests are planned as well~\footnote{\emph{Acknowledgment:} We thank Shrey Baheti from the DREAM:Lab for help with the experiments. This work was supported by grants from VMWare, Microsoft Azure and the Indo-US Science and Technology Forum (IUSSTF).}. 

\bibliographystyle{IEEEtran}
\bibliography{arxiv}
	
\end{document}